\documentclass[eqsecnum,showpacs,pre,twocolumn]{revtex4}

\usepackage{graphicx}
\usepackage{epsfig}
\usepackage{dcolumn}
\usepackage{amsfonts}
\usepackage{amsmath}
\usepackage{amssymb}
\usepackage{bm}

\begin{document}

\newcommand{\brm}[1]{\bm{{\rm #1}}}

\hyphenation{Son-der-forsch-ungs-be-reich}

\title{Logarithmic corrections to scaling in critical percolation and random resistor networks}

\author{Olaf Stenull}
\affiliation{
Department of Physics and Astronomy,
University of Pennsylvania,
Philadelphia, PA 19104,
USA
}

\author{Hans-Karl Janssen}
\affiliation{
Institut f\"{u}r Theoretische Physik III,
Heinrich-Heine-Universit\"{a}t,
Universit\"{a}tsstra{\ss}e 1,
40225 D\"{u}sseldorf,
Germany
}

\date{\today} 

\begin{abstract}
\noindent
We study the critical behavior of various geometrical and transport properties of percolation in 6 dimensions. By employing field theory and renormalization group methods we analyze fluctuation induced logarithmic corrections to scaling up to and including the next to leading correction. Our study comprehends the percolation correlation function, i.e., the probability that 2 given points are connected, and some of the fractal masses describing percolation clusters. To be specific, we calculate the mass of the backbone, the red bonds and the shortest path. Moreover, we study key transport properties of percolation as represented by the random resistor network. We investigate the average 2-point resistance as well as the entire family of multifractal moments of the current distribution.   
\end{abstract}

\maketitle

\section{Introduction}
\label{intro}
\noindent
Percolation~\cite{bunde_havlin_91_etc} is perhaps the most fundamental model for disordered systems. Despite its simplicity percolation has an abundance of applications~\cite{zallen_83}.  It is one of the best studied problems in statistical physics. After decades of enormous interest it is justified to call percolation a major theme of statistical physics. Notwithstanding the intensive work accomplished to date, percolation remains a vivid area of research.

There has been enormous progress in understanding percolation in low dimensions. At small $d$ many aspects of percolation can be studied with reasonable effort and high precision by numerical means. In $d=2$, moreover, several features of percolation are known exactly due to conformal invariance~\cite{cardy_02}. Higher dimensions benefit from being accessible by renormalization group methods. In particular, various scaling exponents describing critical percolation have been calculated using expansions in the deviation $\varepsilon$ from the upper critical dimension 6. In comparison, logarithmic corrections, that are important for the critical behavior at $d=6$, have gained little attention so far. Exceptions can be found in references~\cite{essam&Co_78,aharony_80,ruiz-lorenzo_98} where the leading logarithmic corrections have been analyzed for purely geometric aspects of static percolation such as the percolation probability (probability that a given site belongs to an infinite cluster), the mean-square cluster size and the correlation length. As far as the transport properties of percolation clusters are concerned, logarithmic corrections have not been calculated to date. Part of the reason for this lack of progress was certainly that there were no good numerical estimates available to verify analytical results on logarithmic corrections. In the meantime, however, highly precise numerical results for percolation in high dimensions have become available~\cite{XXX,grassberger_02,osterkamp&Co_2003,grassberger_pers}. Now, there exist Monte Carlo results that clearly indicate the importance of logarithmic corrections~\cite{grassberger_pers}.

In recent years logarithmic corrections observed in simulations on linear polymers have been convincingly explained by field theoretic methods~\cite{GHS94,GHS99}. However, in order to abtain the good agreement between numerics and theory it was necessary to push the analytic results beyond the leading logartihmic correction. It is reasonable to expect that higher order corrections will be likewise important in percolation. The purpose of the present paper is to derive corresponding analytic results for several aspects of the percolation problem with the emphasis on transport properties. 

Our investigation is based on the Harris Lubensky (HL) model~\cite{harris_lubensky_84,harris_kim_lubensky_84,harris_lubensky_87b} for the random resistor network (RRN). In the past, the HL model has proved to be very valuable for studying the transport properties of percolation clusters as well as related problems. It allows to determine the average resistance between two points on a percolation cluster in an elegant way. A generalization of the HL model by Harris~\cite{harris_87} featuring non-linear current-voltage characteristics is suited to study the fractal dimensions of several substructures of percolation clusters. Another generalization by Park, Harris and Lubensky (PHL)~\cite{park_harris_lubensky_87} incorporates the effects of noise and facilitates investigations of the multifractal current distribution on RRNs. We exploit these three models to calculate logarithmic corrections to scaling for the average resistance, the fractal masses of the backbone, the red (singly connected) bonds,  and the chemical (shortest) path as well as of all moments of the current distribution up to and including the next to leading correction.

The percolation models we scrutinize in the present paper belong to one particular type of percolation, viz.\ static isotropic percolation. Nevertheless, logarithmic corrections are expected to be equally important in the critical behavior of other kinds of percolation such as dynamic isotropic percolation and directed percolation. These complimentary topics are/will be  addressed in separate publications~\cite{janssen_stenull_dip_log,janssen_stenull_dp_log}.

The outline of the present paper is as follows:  In Sec.~\ref{theModel} we briefly explain the HL model and its variants. Section~\ref{renormalization} sketches the renormalization of these models and reviews previous results that we need as input as we proceed. The core of our study of logarithmic correction is presented in Section~\ref{logarithmicCorrections}. This section also contains our main results.  Concluding remarks are given in Sec.~\ref{concusions}. In Appendix~\ref{app:amplitudes} we outline our 1-loop calculation of certain amplitudes that enter the logarithmic corrections. Appendix~\ref{schwinger}, finally, contains technical details on the computation of integrals. 

\section{The HL model and its variants}
\label{theModel}
In this section we briefly review the HL model and its generalizations with the aim to provide the reader with background and to establish notation.

\subsection{The RRN}
The HL model is based on earlier ideas by Stephen~\cite{stephen_78}. It represents a field theoretic minimal model for the linear RRN, where  randomly occupied bonds between nearest neighboring sites on a $d$ dimensional lattice are assumed to behave like Ohmic resistors. However, the HL model is not just describing RRN. It also applies to a class of continuous spin systems including the $x$-$y$-model. Here, we are exclusively concerned with its implications for the RRN.

The HL model can be formulated in terms of an order parameter field $\varphi ({\bf x},\vec{\theta})$ which lives on the $d$-dimensional real space with coordinates ${\bf x}$. The variable $\vec{\theta}$ denotes a $D$-fold replicated voltage. For regularization purposes, $\vec{\theta}=\vec{\nu}\Delta \theta $ takes discrete values on a $D$-dimensional torus, the replica space, i.e., $\vec{\nu}$ is chosen to be a $D$-dimensional vector with integer components $\nu^{(\alpha )}$ satisfying $-M<\nu^{(\alpha )}\leq M$ and $\nu^{(\alpha )} = \nu^{(\alpha )} \mod (2M)$. The order parameter field is restricted by the condition $\sum_{\vec{\theta}} \varphi({\bf x},\vec{\theta})=0$. It follows that the replica space Fourier transform $\psi ({\bf x},\vec{\lambda})$ of the order parameter field, defined by 
\begin{eqnarray}
\varphi ({\bf x},\vec{\theta})= (2M)^{-D}\sum_{\vec{\lambda}} \psi ({\bf x},\vec{\lambda}) \exp ( i\vec{\lambda}\cdot \vec{\theta} ) \, ,
\end{eqnarray}
satisfies $\psi ({\bf x},\vec{\lambda} = \vec{0}) =0$. To retrieve physical quantities from the replica formulation one has to study the limit $D\rightarrow 0$, $M\rightarrow \infty $ with $(2M)^{D}\rightarrow 1$ and $\Delta \theta =\theta_{0}/\sqrt{M} \to 0$. Here $\theta_{0}$ is a constant which sets the width of the voltage interval such that $[-\theta_{0} \sqrt{M} < \theta^{(\alpha )} \leq \theta_{0} \sqrt{M}]$. In the limit $D\rightarrow 0$, $M\rightarrow \infty$ the constant $\theta _{0}$ plays the role of a redundant scaling parameter, i.e., the theory is independent of its value.

In terms of $\varphi ({\bf x},\vec{\theta})$ the HL Hamiltonian reads
\begin{eqnarray}
\label{hamiltonian}
{\mathcal{H}}
&=&\int d^{d}x\,\sum_{\vec{\theta}}\bigg\{ \frac{\tau }{2}\varphi({\bf x},\vec{\theta})^{2}+\frac{1}{2}\left[ \nabla \varphi({\bf x},\vec{\theta}) \right]^{2}
\nonumber \\
&+&\frac{w}{2}\left[ \nabla_{\theta } \varphi ({\bf x},\vec{\theta})\right]^{2} + \frac{g}{6} \varphi({\bf x},\vec{\theta})^{3}\bigg\} \, .  
\end{eqnarray}
The parameter $\tau - \tau_c \sim (p_{c}-p)$ specifies the deviation of the occupation probability $p$ from its critical value $p_{c}$. In mean field theory the percolation transition occurs at $\tau = \tau_c =0$. $w$ is proportional to the resistance of the individual random bonds. For $w\rightarrow 0$, ${\mathcal{H}}$ reduces to the Hamiltonian for the $n=(2M)^{D}$-state Potts model~\cite{Zia_Wallace_75} with $n\rightarrow 1$ for $D\rightarrow 0$. Hence, ${\mathcal{H}}$ describes purely geometrical percolation in this limit~\cite{kasteleyn_fortuin_69}. 

The 2-point correlation function
\begin{eqnarray}
\label{defCorr}
G_2 ( \brm{x}, \brm{x}^\prime , \vec{\lambda}  ) = \left\langle \psi ( \brm{x},\vec{\lambda}) \psi (\brm{x}^\prime, - \vec{\lambda}) \right\rangle
\end{eqnarray}
has the valuable property of being a generating function for the average two-point resistance. This feature can be understood by noting that
\begin{eqnarray}
\label{genFkt}
G_2 ( \brm{x}, \brm{x}^\prime , \vec{\lambda} )
&=&\bigg\langle {\rm \exp }\bigg(-
\frac{\vec{\lambda}^{2}}{2}R (\brm{x}, \brm{x}^\prime ) \bigg)\bigg\rangle_C 
\nonumber \\
&=&\bigg\langle \chi (\brm{x}, \brm{x}^\prime ) \, \exp \bigg(-\frac{\vec{\lambda}^{2}}{2}  R (\brm{x}, \brm{x}^\prime )\bigg)\bigg\rangle_C
\nonumber \\
&=& P  (\brm{x}, \brm{x}^\prime )\bigg\langle {\rm \exp }\bigg( -\frac{\vec{\lambda}^{2}}{2} R (\brm{x}, \brm{x}^\prime )\bigg)\bigg\rangle_C^\prime \, .  
\end{eqnarray}
Here $R (\brm{x}, \brm{x}^\prime )$ is the total  resistance between two arbitrary points ${\bf x}$ and $\brm{x}^\prime$.  $\chi (\brm{x}, \brm{x}^\prime )$ is an indicator function which is one if ${\bf x}$ is connected to $\brm{x}^\prime$, and zero otherwise. $\langle \cdots \rangle_C$ denotes the disorder average over all configurations of the diluted lattice. $\langle \cdots \rangle_C^\prime$ stands for disorder averaging conditional to the constraint that ${\bf x}$ and $\brm{x}^\prime$ are connected. $P (\brm{x}, \brm{x}^\prime ) = \langle \chi  (\brm{x}, \brm{x}^\prime ) \rangle_C$ is nothing more than the correlation function for usual (purely geometric) percolation, i.e., the probability for ${\bf x}$ and $\brm{x}^\prime$ being connected. From Eq.~(\ref{genFkt}) it follows that one can extract the average resistance
\begin{eqnarray}
M_R  (\brm{x}, \brm{x}^\prime )  = \langle  R (\brm{x}, \brm{x}^\prime )  \rangle^\prime_C
\end{eqnarray}
essentially by taking the derivative of $G ( \brm{x}, \brm{x}^\prime , \vec{\lambda})$ with respect to $\vec{\lambda}^{2}$.

\subsection{The nonlinear RRN}
In Ref.~\cite{harris_87} Harris implemented ideas by Kenkel and Straley~\cite{kenkel_straley_82} and generalized the HL model so that it captures nonlinear voltage-current characteristics of the type $V \sim I^r$. The Hamiltonian for the nonlinear RRN is given by Eq.~(\ref{hamiltonian}) with the replacement
\begin{eqnarray}
\label{replace}
&&\frac{w}{2}\left[ \nabla_{\theta } \varphi ({\bf x},\vec{\theta})\right]^{2} \to
\nonumber \\
&&\ \, - \frac{w_r}{2} \varphi ( 
{\rm{\bf x}} , \vec{\theta} ) \sum_{\alpha =1}^D \left( - \frac{\partial}{\partial 
\theta^{(\alpha )}} \right)^{r+1} \varphi ( {\rm{\bf x}} , \vec{\theta}) \, ,  
\end{eqnarray}
where $w_r$ is proportional to the nonlinear bond resistance. The 2-point correlation function in the nonlinear model satisfies
\begin{eqnarray}
\label{genFktnonlin}
G_2 ( \brm{x}, \brm{x}^\prime , \vec{\lambda} )
= P  (\brm{x}, \brm{x}^\prime )\bigg\langle {\rm \exp }\bigg( \frac{\Lambda_r (\vec{\lambda})}{r+1} R_r (\brm{x}, \brm{x}^\prime )\bigg)\bigg\rangle_C^\prime \, , 
\end{eqnarray}
where $\Lambda_r (\vec{\lambda}) = \sum_{\alpha =1}^D (-i \lambda^{(\alpha )} )^{r+1}$ and $R_r (\brm{x}, \brm{x}^\prime )$ is the total nonlinear resistance between the two points ${\bf x}$ and $\brm{x}^\prime$. To obtain the average nonlinear resistance
\begin{eqnarray}
M_{R_r}  (\brm{x}, \brm{x}^\prime )  = \langle  R_r (\brm{x}, \brm{x}^\prime )  \rangle^\prime_C
\end{eqnarray}
one just needs to take, apart from factors, the derivative of Eq.~(\ref{genFktnonlin}) with respect to $\Lambda_r (\vec{\lambda})$. For $r\to 1$ one retrieves, of course, the linear average resistance $M_R (\brm{x}, \brm{x}^\prime)$.

Generalizing the RRN to the nonlinear case means more than just an academic exercise. The great value of the nonlinear RRN is that it can be used to map out different fractal substructures of percolation clusters. Looking at the overall dissipated power, it is not difficult to see that  
\begin{eqnarray}
\label{relDB}
\lim_{r \to -1^+} M_{R_r}  (\brm{x}, \brm{x}^\prime )  \sim M_B \, ,
\end{eqnarray}
where $M_B$ is the average number of bonds (the mass) of the backbone. Moreover, it has been shown by Blumenfeld and Aharony~\cite{blumenfeld_aharony_85} that
\begin{eqnarray}
\lim_{r \to \infty} M_{R_r}  (\brm{x}, \brm{x}^\prime )  \sim M_{\text{red}} \, ,
\end{eqnarray}
where $M_{\text{red}}$ stands for the mass of the red bonds and
\begin{eqnarray}
\label{relMin}
\lim_{r \to 0^+} M_{R_r}  (\brm{x}, \brm{x}^\prime )  \sim M_{\text{min}} \, ,
\end{eqnarray}
where $M_{\text{min}}$ is the mass of the chemical path.

\subsection{The noisy RRN}
PHL studied a RRN with microscopic noise in the sense, that the conductances of the individual occupied bonds are drawn from a probability distribution $f$ (e.g., a suitable Gaussian). In order to perform both, the average $\langle \cdots \rangle_C$ over the diluted lattice configurations and the noise average $\{ \cdots \}_f$, PHL introduced ($D\times E$)-fold replicated voltages $\tensor{\theta}=\tensor{\nu}\Delta \theta $ living on a ($D\times E$)-dimensional torus. That means $\tensor{\nu}$ is chosen to be a matrix with integer components $\nu^{(\alpha , \beta)}$ satisfying $-M<\nu^{(\alpha , \beta )}\leq M$ and $\nu^{(\alpha ,\beta)} = \nu^{(\alpha ,\beta)} \mod (2M)$.

The Hamiltonian introduced by PHL can be cast as 
\begin{eqnarray}
\label{PHLhamiltonian}
{\mathcal{H}}
&=&\int d^{d}x\,\sum_{\tensor{\theta}}\bigg\{ \frac{\tau }{2}\varphi({\bf x},\tensor{\theta})^{2}+\frac{1}{2}\left[ \nabla \varphi({\bf x},\tensor{\theta}) \right]^{2}
\nonumber \\
&-&\frac{w}{2} \,  \varphi({\bf x},\tensor{\theta})  \sum_{\alpha , \beta =1}^{D,E} \frac{\partial^2}{\left( \partial 
\theta^{(\alpha ,\beta 
)} \right)^2 }  \,  \varphi({\bf x},\tensor{\theta})
\nonumber \\
&+&\varphi({\bf x},\tensor{\theta})  \sum_{l=2}^\infty v_l \sum_{\beta=1}^E \left[ \sum_{\alpha =1}^D   
\frac{- \partial^2}{\left( \partial \theta^{(\alpha ,\beta )} \right)^2 } \right]^{l} \varphi({\bf x},\tensor{\theta})
\nonumber \\
&+& \frac{g}{6} \, \varphi({\bf x},\tensor{\theta})^{3}\bigg\} \, ,  
\end{eqnarray}
with $v_l$ being proportional to the $l$th cumulant of the distribution $f$. In contrast to the relevant $w$, the $v_l$ represent dangerous irrelevant couplings. The consequences of their dangerous irrelevance will be explained below.

The 2-point correlation function $G_2 ( \brm{x}, \brm{x}^\prime , \tensor{\lambda})$ generalizing (\ref{defCorr}) has the property
\begin{eqnarray}
\label{cumulantGenFkt}
G_2 ( \brm{x}, \brm{x}^\prime, \tensor{\lambda} ) = 
\bigg\langle  \exp \bigg[ \sum_{l=1}^\infty \frac{(-1)^l}{2^l \, l!} K_l ( 
\tensor{\lambda} ) \{ R \left( x,x^\prime \right)^l \}_f^{(c)} \bigg]  \bigg\rangle_C 
\nonumber \\
\end{eqnarray}
where $\{ R \left( \brm{x}, \brm{x}^\prime \right)^l \}_f^{(c)}$ is the $l$th cumulant of the total resistance $R ( \brm{x}, \brm{x}^\prime)$ with respect to the distribution $f$ and $K_l$ is defined by
\begin{eqnarray}
\label{defOfK_l}
K_l ( \tensor{\lambda}) = \sum_{\beta =1}^E \bigg[ \sum_{\alpha =1}^D \left( \lambda^{(\alpha ,\beta )} \right)^2 \bigg]^l \, .
\end{eqnarray}
In other words, $G ( \brm{x}, \brm{x}^\prime , \tensor{\lambda})$ is a generating function for the noise cumulant
\begin{eqnarray}
\label{defCumulant}
C_R^{(l)}(\brm{x}, \brm{x}^\prime) = \left\langle  \{ R (\brm{x}, \brm{x}^\prime )^l \}_f^{(c)} \right\rangle_C^\prime  \, .
\end{eqnarray}
Though the noise cumulants are certainly interesting in their own right, our main motivation to study them is that they are closely related to the multifractal moments 
\begin{eqnarray}
\label{multiMoment}
M_I^{(l)}(\brm{x}, \brm{x}^\prime) = \Big\langle \sum_b \left( I_b/I \right)^{2l} \Big\rangle_C^\prime 
\end{eqnarray}
of the current distribution. Here, $I$ denotes an external current inserted at $\brm{x}$ and withdrawn at $\brm{x}^\prime$, $I_b$ is the microscopic current flowing through bond $b$ and the sum runs over all bonds on the cluster connecting  $\brm{x}$ and $\brm{x}^\prime$. By virtue of Cohn's theorem~\cite{cohn_50}, the noise cumulants and the multifractal moments are related by
\begin{eqnarray}
\label{finalCumulant}
M_I^{(l)}(\brm{x}, \brm{x}^\prime) \sim C_R^{(l)}(\brm{x}, \brm{x}^\prime) \, . 
\end{eqnarray}

\section{Renormalization and scaling}
\label{renormalization}
In this section we review the previous results of our renormalization group analysis~\cite{stenull_2000,stenull_janssen_oerding_99,janssen_stenull_oerding_99,janssen_stenull_99,stenull_janssen_epl_2000,stenull_janssen_2001} for the 3 models explained in the preceding section. We establish intermediate results that are required for deriving the logarithmic corrections we are interested in. For the sake of briefness, we review the linear and the nonlinear case in one go. This represents no difficulty, since the linear case can be retrieved from the nonlinear case simply by taking the limit $r\to 1$. We present the noisy RRN separately, because the dangerous irrelevance of the $v_l$ brings about some intricacies that are absent in the linear and nonlinear RRN. For background on the methods used in the remainder we refer to~\cite{amit_zinn-justin}.

\subsection{The linear/nonlinear RRN}
A central stage in the renormalization group analysis of the RRN is, as usual, a diagrammatic perturbation calculation. The ultraviolet (UV) divergences encountered in computing the diagrams can be handled by dimensional regularization. In dimensional regularization the UV divergences appear as poles in the deviation $\varepsilon = 6-d$ from the upper critical dimension 6 for the RRN. These poles can be eliminated by employing the renormalization scheme
\begin{subequations}
\label{renScheme}
\begin{eqnarray}
\varphi \to \mathring{\varphi} &=& Z^{1/2} \varphi \, ,
\\
\tau \to \mathring{\tau} &=& Z^{-1} Z_{\tau} \tau  \, ,
\\
w_r \to \mathring{w}_r &=& Z^{-1} Z_{w_r} w_r \, , 
\\
g \to \mathring{g} &=& Z^{-3/2} Z_u^{1/2} G_\varepsilon^{-1/2} u^{1/2} \, \mu^{\varepsilon/2} \, ,
\end{eqnarray}
\end{subequations}
where the $\mathring{}$ indicates unrenormalized quantities. The factor $G_\varepsilon = (4\pi )^{-d/2}\Gamma (1 + \varepsilon /2)$ is introduced for later convenience. $Z$, $Z_\tau$, and $Z_u$ are the usual percolation $Z$ factors known to three-loop order~\cite{alcantara_80}. 

The renormalization factor $Z_{w_r}$ can be calculated in an elegant and efficient way by utilizing our real-world interpretation~\cite{stenull_janssen_oerding_99} of the Feynman diagrams for the RRN. For arbitrary $r$ $Z_{w_r}$ is known to 1-loop order. However, we are less interested in the most general case than in those $r$ that have a clear physical significance, c.f.\ Sec.~\ref{theModel}. In Rev.~\cite{stenull_janssen_oerding_99} we have calculated $Z_w = Z_{w_1}$ for the linear RRN to 2-loop order. In our work on the nonlinear RRN~\cite{janssen_stenull_oerding_99,janssen_stenull_99}, we computed $Z_0 = \lim_{r\to 0^+} Z_{w_r}$ to 2-loop order and $Z_{-1} = \lim_{r\to -1^+} Z_{w_r}$ to 3-loop order. Moreover, we showed explicitly to 3-loop order, that $Z_\infty = \lim_{r\to \infty} Z_{w_r} = Z_\tau$ as had to be expected from rigorous results by Coniglio~\cite{coniglio_81_82}. 

The critical behavior of any connected $N$-point correlation function of the order parameter field is governed by an Gell-Mann--Low renormalization group equation (RGE). In the remainder we will use  two equitable types of notation for the $N$-point functions, depending on which is beneficial to the actual argument, viz.\ $G_N (\{ {\rm{\bf x}} ,w_r \Lambda_r (\vec{\lambda}) \} ; u, \tau, \mu )$ and $G_N (\{ {\rm{\bf x}} ,\vec{\lambda} \} ; u, \tau, w_r, \mu )$. Our RGE reads
\begin{eqnarray}
\label{rge}
&&\left[ \mu \frac{\partial }{\partial \mu} + \beta \frac{\partial }{\partial u} + \tau \kappa \frac{\partial }{\partial \tau} + w_r \zeta_r 
\frac{\partial }{\partial w_r} + \frac{N}{2} \gamma \right] 
\nonumber \\
&&\times \, 
G_N \big(\big\{ {\rm{\bf x}} ,w_r \Lambda_r (\vec{\lambda}) \big\} ; u, \tau, \mu \big) = 0 \, .
\end{eqnarray}
The Wilson functions appearing in the RGE~(\ref{rge}) are given to 2-loop order by
\begin{subequations}
\label{wilson2loop}
\begin{eqnarray}
\label{wilsonGammaRes}
\gamma \left( u \right) &=& - \frac{1}{6}\, u + \frac{37}{216}\, u^2 + O \left( u^3 \right) \, ,
\\
\kappa \left( u \right) &=&   \frac{5}{6}\, u - \frac{193}{108}\, u^2 + O \left( u^3 \right) \, ,
\\
\label{wilsonZetaRes}
\zeta_r \left( u \right) &=&    \zeta_{r,1} \, u + \zeta_{r,2} \, u^2 + O \left( u^3 \right) \, ,
\\
\label{betauRes}
\beta \left( u \right) &=&  - \varepsilon \, u + \frac{7}{2}\, u^2 - \frac{671}{72}\, u^3 
\nonumber \\
&+& \left(  \frac{414031}{10368} + \frac{93 \, \zeta (3)}{4} \right) u^4 + O \left( u^5 \right) \, .
\end{eqnarray}
\end{subequations}
The $\zeta$ in Eq.~(\ref{betauRes}) stands for the Riemann $\zeta$ function and should not be confused with the $\zeta$'s featured in (\ref{wilsonZetaRes}). The values of $\zeta_{r,1}$ and $\zeta_{r,2}$ are given in Table~\ref{tab:coeffs}. Note that we have displayed $\beta$ up to 3-loop order since the accuracy of the sort of field theoretic prediction we have in mind depends noticeably on a good knowledge of $\beta$. In the following we will use for the Wilson functions an abbreviated notation of the type $f(u)=f_{1}u+f_{2}u^{2}+\cdots$. For example, we will write Eq.~(\ref{betauRes}) as $\beta (u) = \beta_1 u + \beta_2 u^2 + \beta_3 u^3 + \beta_4 u^4 + O(u^5)$ and likewise for the other Wilson functions.
\begin{table}
\caption{The coefficients $\zeta_{r,1}$ and $\zeta_{r,2}$ appearing in Eq.~(\ref{wilsonZetaRes}).}
\label{tab:coeffs}
\begin{tabular}{c||c|c|c|c}
\hline \hline
$\quad r \quad $ & $-1$ & $0$ & $1$ & $\infty$ \\ \hline
$ \zeta_{r,1}$ & -$\frac{1}{6}$ & $\frac{7}{12}$ & $\frac{2}{3}$ & $\frac{5}{6}$ \\
\hline
$ \zeta_{r,2} $ & $\frac{145}{216}$ & $-\frac{1}{32} \left(  \frac{1747}{54}  + 9 \ln (3) - 10 \ln (2) \right)$ & $-\frac{47}{36}$ & $ - \frac{193}{108}$ \\
\hline \hline
\end{tabular}
\end{table}

The RGE can be solved in terms of a single flow parameter $\ell$ by introducing the characteristics
\begin{subequations}
\begin{eqnarray}
\ell \frac{\partial \bar{\mu}(\ell)}{\partial l} &=& \bar{\mu} (\ell) \, , \quad \bar{\mu}(1)=\mu \ ,
 \\
\label{charBeta}
\ell \frac{\partial \bar{u}(\ell)}{\partial \ell} &=& \beta \left( \bar{u}(\ell) \right) \, , \quad \bar{u}(1)=u \, ,
 \\
\label{charTau}
\ell \frac{\partial}{\partial \ell} \ln \bar{\tau} \left( \bar{u}(\ell) \right)&=& \kappa \left( \bar{u}(\ell) \right) \, , \quad \bar{\tau}(u)=\tau \, ,
 \\
\ell \frac{\partial}{\partial \ell} \ln \bar{w}_r \left( \bar{u}(\ell) \right)&=& \zeta_r \left( \bar{u}(\ell) \right) \, , \quad \bar{w}_r(u)=w_r \, ,
 \\
\label{charZ}
\ell \frac{\partial}{\partial \ell} \ln \bar{Z} \left( \bar{u}(\ell) \right)&=& \gamma \left( \bar{u}(\ell) \right) \, , \quad \bar{Z}(u)=1 \, .
\end{eqnarray}
\end{subequations}
These characteristics describe how the parameters transform if we change the momentum scale $\mu $ according to $\mu \to \bar{\mu}(\ell)=\mu \ell$. Supplementing our solution to the RGE with a dimensional analysis (to account for naive dimensions) we find 
\begin{eqnarray}
\label{RGEsol}
&&G_N \big(\big\{ {\rm{\bf x}} ,w_r \Lambda_r (\vec{\lambda}) \big\} ;  u, \tau , \mu \big) = 
\ell^{(d-2)N/2} \bar{Z} (\bar{u})^{N/2}
\nonumber \\
&& \times \,
G_N \big( \big\{ \ell{\rm{\bf x}} ,\ell^{-2} \bar{w}_r ( \bar{u}) \Lambda_r (\vec{\lambda}) \big\} ;  \bar{u}, \ell^{-2}  \bar{\tau} ( \bar{u}), \mu \big) \, .
\end{eqnarray}
Of course, Eq.~(\ref{RGEsol}) is, as it stands, only of formal value. The functions $ \bar{u}$, $\bar{Z} (\bar{u})$ and so on have to be filled with life. This will be done in Sec.~\ref{logarithmicCorrections}.

To gain information on the observables of interest, we have to take a closer look at $N=2$. Moreover, we have to make an appropriate choice for the flow parameter $\ell$. Since we are interested in criticality, $\tau =0$, and long length scales, we choose
\begin{eqnarray}
\label{choicel}
\ell = \frac{2 \, X_0}{\mu | \brm{x} -  \brm{x}^\prime|} \, ,
\end{eqnarray}
where $X_0$ is a constant of the order of unity. In the following we set $\brm{x}^\prime =   \brm{0}$ for notational simplicity. At $d =6$ dimensions we then get
\begin{eqnarray}
\label{struct}
&&G_2 \big(  \brm{x} ,w_r \Lambda_r (\vec{\lambda})  ;  u, 0 , \mu \big) = 
\nonumber \\
&& \times \, \left( \frac{| \brm{x}|}{2 \, X_0} \right)^{-4} \bar{Z} ( \bar{u}) \, 
 \bigg\{ 
G_2 \big( 2 X_0 , 0 ;  \bar{u}, 0, 1 \big) 
\nonumber \\
&& + \, \left( \frac{| \brm{x}|}{2 \, X_0} \right)^{2} \bar{w}_r ( \bar{u}) \Lambda_r (\vec{\lambda}) \, G^\prime_2 \big( 2 X_0 , 0 ;  \bar{u}, 0, 1 \big) 
\nonumber \\
&& + \, \cdots \bigg\} \, ,
\end{eqnarray}
where $G^\prime_2 = \partial G_2 / \partial \bar{w}_r \Lambda_r$. The scaling functions $G_2$ and $G^\prime_2$ have the loop expansions
\begin{subequations}
\begin{eqnarray}
\label{kaffeedurst}
G_2 \big( 2 X_0 , 0 ;  \bar{u}, 0, 1 \big) &=& G_2^{(0)} \left\{  1  + A_P \left( X_0 \right) \, \bar{u}+ O (\bar{u}^2) \right\} \, ,
\nonumber \\
\\
G_2^\prime \big( 2 X_0 , 0 ;  \bar{u}, 0, 1 \big) &=& G_2^{\prime (0)} \left\{  1  + A_{w_r} \left( X_0 \right) \bar{u} + O (\bar{u}^2) \right\} \, ,
\nonumber \\
\end{eqnarray}
\end{subequations}
where $G_2^{(0)}$ and $G_2^{\prime (0)}$ denote the respective 0-loop contributions. $A_P$ and $A_{w_r}$ are amplitudes that, to our knowledge, have not been calculated hitherto. Unlike critical exponents these amplitudes are not entirely determined by the renormalization mapping itself. On the diagrammatic level this means that we cannot restrict ourself to consider the UV divergent parts of the Feynman diagrams. Rather, we need to include regular parts of the diagrams, i.e., those not associated with $\varepsilon$ poles. Appendix~\ref{app:amplitudes} outlines these calculations that we carried out to 1-loop order.

Now we are in the position to extract the structure of our observables of interest. For the usual percolation correlation function we deduce from Eq.~(\ref{struct}) and (\ref{kaffeedurst}) that
\begin{eqnarray}
\label{structP}
&&P (\brm{x}) = G_2 \big(  \brm{x} ,0  ;  u, 0 , \mu \big) 
\nonumber \\
&&\sim \, \left( \frac{| \brm{x}|}{2 \, X_0} \right)^{-4}  \bar{Z} ( \bar{u})  \,  \left\{  1  + A_P \left( X_0 \right) \, \bar{u}+ O (\bar{u}^2) \right\} \, ,
\end{eqnarray}
Equation~(\ref{genFktnonlin}) in conjunction with (\ref{struct}) and (\ref{kaffeedurst}) gives for the average nonlinear resistance
\begin{eqnarray}
\label{structMr}
M_{R_r} (\brm{x}) \sim \left( \frac{| \brm{x}|}{2 \, X_0} \right)^{2}  \bar{w}_r ( \bar{u}) \,  
 \left\{  1  + A_{R_r} \left( X_0 \right) \, \bar{u}+ O (\bar{u}^2) \right\} \, ,
\nonumber \\
\end{eqnarray}
where we introduced the amplitudes
\begin{eqnarray}
\label{defARr}
A_{R_r} \left( X_0 \right) = A_{w_r} \left( X_0 \right) - A_P \left( X_0 \right) \, .
\end{eqnarray}
To 1-loop order, see Appendix~\ref{app:amplitudes}, the amplitudes appearing in Eqs.~(\ref{structP}) and (\ref{structMr}) are given by
\begin{subequations}
\begin{eqnarray}
\label{resAP}
A_P (X_0) &=&  \frac{5}{36} + \frac{1}{6} \mathcal{Z} (X_0) \, ,
\\
\label{resAR1}
A_{R_1} (X_0) &=&  - \frac{11}{36} - \frac{2}{3} \mathcal{Z} (X_0) \, ,
\\
\label{resAR-1}
A_{R_{-1}} (X_0) &=&   \frac{5}{36} + \frac{1}{6} \mathcal{Z} (X_0) \, ,
\\
\label{resAred}
A_{R_{\infty}} (X_0) &=&   - \frac{13}{36} - \frac{5}{6} \mathcal{Z} (X_0) \, ,
\\
\label{resAmin}
A_{R_{0}} (X_0) &=&   - \frac{1}{9} - \frac{1}{4} \ln 2 - \frac{7}{12} \mathcal{Z} (X_0) \, .
\end{eqnarray}
\end{subequations}
Here we used the shorthand $\mathcal{Z} (X_0) = \gamma + \ln X_0$ with $\gamma = 0.577215...$ being Euler's constant.

\subsection{The noisy RRN}
Since the $v_l$ are irrelevant, they cannot be treated in the same manner as the relevant $w_r$. Such an attempt would poison the perturbation expansion. Properly, the $v_l$ can be treated via  insertions of the operator
\begin{eqnarray}
{\mathcal{O}}^{(l)} = - \frac{1}{2} v_l \int d^d p \, \sum_{\tensor{\lambda}} K_l ( \tensor{\lambda} ) \phi ( {\rm{\bf p}} , \tensor{\lambda} ) \phi ( - {\rm{\bf p}} , - \tensor{\lambda} ) \, ,
\end{eqnarray}
where $\phi ( {\rm{\bf p}} , \tensor{\lambda} )$ denotes the Fourier transform of $\varphi ( {\rm{\bf x}} , \tensor{\theta} )$. Due to its irrelevance, these insertions generate a multitude of terms corresponding to operators with equal or lower naive dimension than ${\mathcal{O}}^{(l)}$. All these operators have to be taken into account in the renormalization process. The operators of lower naive dimension, however, merely lead to subdominant corrections and can be ignored for our purposes. Keeping all the operators of the same naive dimension, we have a renormalization in matrix form
\begin{eqnarray}
\underline{{\mathcal{O}}}^{(l)} \to \underline{{\mathring {\mathcal{O}}}}^{(l)} = \underline{\underline{Z}}^{(l)} \underline{{\mathcal{O}}}^{(l)} \, .
\end{eqnarray}
The vector $\underline{{\mathcal{O}}}^{(l)} = ( {\mathcal{O}}^{(l)}, {\mathcal{O}}_2^{(l)}, \cdots )$ contains the family associated with ${\mathcal{O}}^{(l)}$. For the remaining renormalizations, we employ the scheme~(\ref{renScheme}). The ${\mathcal{O}}^{(l)}$ have the feature that they are master operators. For details on the notion of master and slave operators we refer the reader to~\cite{stenull_janssen_epl_2000,stenull_janssen_2001}. The master operator property has the  important consequence that the renormalization matrix $\underline{\underline{Z}}^{(l)}$ has a simple structure, 
\begin{eqnarray}
\underline{\underline{Z}}^{(l)} = 
\left(
\begin{array}{cccc}
Z^{(l)} & \Diamond & \cdots & \Diamond \\
0       & \Diamond & \cdots & \Diamond \\
\vdots  & \vdots & \ddots & \vdots \\
0       & \Diamond & \cdots & \Diamond
\end{array}
\right) \, . 
\end{eqnarray}
$\Diamond$ stands for elements that we do not need evaluate. 

The connected $N$-point correlation functions with an insertion of $\underline{\mathcal{O}}^{(l)}$ are governed by the RGE
\begin{eqnarray}
\label{noisyRGG}
&& \left\{ \left[ \mu \frac{\partial }{\partial \mu} + \beta \frac{\partial 
}{\partial u} + \tau \kappa \frac{\partial }{\partial \tau} + w \zeta \frac{\partial 
}{\partial w} + \frac{N}{2} \gamma 
\right] \underline{\underline{1}} + \underline{\underline{\gamma}}^{(l)} \right\} 
\nonumber \\
&& \times \, 
G_N \big( \big\{ {\rm{\bf x}} , w \tensor{\lambda}^2 \big\} ; u, \tau, \mu \big)_{\underline{\mathcal{O}}^{(l)}} = 0 \, 
\end{eqnarray}
where $\underline{\underline{1}}$ is a unit matrix and where
\begin{eqnarray}
\label{noisyWilson}
\underline{\underline{\gamma}}^{(l)} \left( u \right) &=& - \mu \frac{\partial 
}{\partial \mu} \ln \underline{\underline{Z}}^{(l)}  \bigg|_0 \, .
\end{eqnarray}
The only element of $\underline{\underline{\gamma}}^{(l)}$ that we need for our purposes is
\begin{eqnarray}
\label{multiGamma}
\gamma^{(l)} = \gamma_1^{(l)} u + \gamma_2^{(l)} u^2 \, ,
\end{eqnarray}
where we have used the shorthands
\begin{subequations}
\label{gl}
\begin{eqnarray}
\gamma_1^{(l)} = \frac{-1 + 15\,l + 10\,l^2}
  {6\,\left( 1 + l \right) \,\left( 1 + 2\,l \right) }
\end{eqnarray}
and
\begin{widetext}
\begin{eqnarray}
\gamma_2^{(l)} =  \frac{145 - l\,\left( 909 + 2\,l\,
        \left( 4713 + l\,\left( 11727 + 
             2\,l\,\left( 6459 + 386\,l\,\left( 9 + 2\,l \right)  \right) 
             \right)  \right)  \right)  + 
    288\,{\left( 1 + l \right) }^2\,{\left( 1 + 2\,l \right) }^2\, H(2l)}{216\,{\left( 1 + l \right) }^3\,
    {\left( 1 + 2\,l \right) }^3} \, ,
\end{eqnarray}
\end{widetext}
\end{subequations}
where $H(n) =  \sum_{k=1}^n 1/k$ is a harmonic number. 

The RGE~(\ref{noisyRGG}) can be solved via Introducing the function $\bar{\underline{\underline{Z}}}^{(l)} (\bar{u})$ governed by the characteristic
\begin{eqnarray}
\label{multichar}
\ell \frac{\partial}{\partial \ell} \ln \bar{\underline{\underline{Z}}}^{(l)} \left( \bar{u}(\ell ) \right)&=&- \underline{\underline{\gamma}}^{(l)}  \left( \bar{u}(\ell ) \right) \, , \quad \bar{\underline{\underline{Z}}}(u)= \underline{\underline{1}}  \, .
\end{eqnarray}
From the fixed point solution we derive
\begin{eqnarray}
\label{RGEsolNoisy3}
&&G_N \big( \big\{ {\rm{\bf x}} ,w \tensor{\lambda}^2 \big\} ;  u, \tau , \mu \big)_{\mathcal{A}^{(l)}} = 
\ell^{(d-2)N/2 -2}\bar{Z} (\bar{u})^{N/2} \bar{Z}^{(l)} (\bar{u})
\nonumber \\
&& \times \,
G_N \big( \big\{ \ell {\rm{\bf x}} , \ell^{-2} \bar{w} (\bar{u}) \tensor{\lambda}^2 \big\} ;  \bar{u},   \ell^{-2} \bar{\tau} (\bar{u}), \mu \big)_{\mathcal{A}^{(l)}}  \, ,
\end{eqnarray}
where 
\begin{eqnarray}
\label{Adef}
\mathcal{A}^{(l)} =  \mathcal{O}^{(l)} + \cdots 
\end{eqnarray}
is an operator whose form is determined by the eigenvectors of the RGE. The ellipsis in Eq.~(\ref{Adef}) stands for various of the other operators generated in the perturbation calculation (slaves, cf.~Refs.~\cite{stenull_janssen_epl_2000,stenull_janssen_2001}).

To extract the multifractal moments, we have to scrutinize the case $N=2$. With our choice for the flow parameter, Eq.~(\ref{choicel}), we can write in 6 dimensions
\begin{eqnarray}
\label{scalemitIns}
&&G_2 \big(  {\rm{\bf x}} ,w \tensor{\lambda}^2  ;  u, 0 , \mu \big)_{\mathcal{A}^{(l)}} = - v_l K_l( \tensor{\lambda} )   \left( \frac{| \brm{x}|}{2 \, X_0} \right)^{-2} 
\nonumber \\
&& \times \, \bar{Z} (\bar{u}) \bar{Z}^{(l)} (\bar{u}) \, F^{(0)}  \left\{  1  + A_{v_l} \left( X_0 \right) \bar{u} + O (\bar{u}^2) \right\} 
\nonumber \\
&& + \cdots \, ,
\end{eqnarray}
where $F^{(0)}$ is a 0-loop scaling function identical to $G_2^{\prime (0)}$ and where $A_{v_l}$ is an amplitude that we calculate in Appendix~\ref{app:amplitudes}. Equations~(\ref{cumulantGenFkt}), (\ref{finalCumulant}), (\ref{struct}), (\ref{kaffeedurst}) and (\ref{scalemitIns}) tell us that the multifractal moments are of the structure
\begin{eqnarray}
\label{structMI}
M_{I}^{(l)} (\brm{x}) \sim \left( \frac{| \brm{x}|}{2 \, X_0} \right)^{2}  \bar{Z}^{(l)} (\bar{u}) \,  
 \left\{  1  + A_{I}^{(l)} \left( X_0 \right) \, \bar{u}+ O (\bar{u}^2) \right\} \, ,
\nonumber \\
\end{eqnarray}
where we have introduced
\begin{eqnarray}
\label{defAIl}
A_{I}^{(l)} \left( X_0 \right) = A_{v_l} \left( X_0 \right) - A_P \left( X_0 \right) \, .
\end{eqnarray}
To 1-loop order, see Appendix~\ref{app:amplitudes}, this amplitude is given by 
\begin{eqnarray}
\label{resAIl}
&&A_I^{(l)} (X_0) =   - \frac{13}{36} - \frac{5}{6} \mathcal{Z} (X_0) + \frac{1}{(2l + 1)(2l+2)} 
 \\ 
&&\times \,  \big[  -1 + 2 \mathcal{Z} (X_0) - \Psi (2l+1) - \Psi (2) + 2\, \Psi (2l +3) \big] \, ,
\nonumber
\end{eqnarray}
where $\Psi$ stands for the Digamma function~\cite{abramowitz_stegun_65}.

\section{Critical Behavior in $d=6$}
\label{logarithmicCorrections}
Having set the stage, we now determine the sought-after logarithmic corrections to the scaling behavior in $d=6$. The basic step that remains to be performed resides in solving the flow equations for the scaling parameters. Once we have these solutions, our final results are readily stated since we already know the amplitudes $A_P \left( X_0 \right)$, $A_{R_r} \left( X_0 \right)$ and $A_{I}^{(l)} \left( X_0 \right)$ from Sec.~\ref{renormalization}. 

\subsection{Solving the characteristics}
Since the characteristics (\ref{charBeta}) to (\ref{charBeta}) all depend on $\bar{u}(\ell)$, we start with solving (\ref{charBeta}). By separation of variables and Taylor expansion we get
\begin{eqnarray}
\label{dglu}
\frac{d \, \ell}{\ell} = \frac{1}{\beta_2} \frac{d \bar{u}}{\bar{u}^2} - \frac{\beta_3}{\beta_2^2} \frac{d \bar{u}}{\bar{u}} + \frac{\beta_3^2 - \beta_2 \beta_4}{\beta_2^3} \, d\bar{u} + O \left( \bar{u} \right) d\bar{u}  \, .
\end{eqnarray}
Therefore, by integrating
\begin{eqnarray}
\label{solu1}
\ln (\ell /\ell_0) &=&  - \frac{1}{\beta_2} \frac{1}{\bar{u}} - \frac{\beta_3}{\beta_2^2} \ln (\bar{u}) + \frac{\beta_3^2 - \beta_2 \beta_4}{\beta_2^3} \, \bar{u} 
\nonumber \\
&+& O \left( \bar{u}^2 \right)   \, ,
\end{eqnarray}
where $\ell_0$ is an integration constant. With our choice for the flow parameter (\ref{choicel}) we obtain 
\begin{eqnarray}
\label{solx}
\frac{|\brm{x}|}{x_0} &=&  \bar{u}^{-a_{|\brm{x}|}} \exp \left( \frac{1}{\beta_2 \bar{u}} +  c_{|\brm{x}|} \,  \bar{u} \right)  \left[ 1+ O \left( \bar{u}^2 \right) \right]  \, ,
\end{eqnarray}
with a nonuniversal constant $x_0 = 2 X_0 /(\mu \ell_0)$ that defines a length scale and the coefficients
\begin{subequations}
\begin{eqnarray}
a_{|\brm{x}|} &=& - \frac{\beta_3}{\beta_2^2} = \frac{671}{882} = 0.76077
\\
\beta_2 &=& \frac{7}{2} = 3.5
\\
c_{|\brm{x}|} &=& \frac{\beta_2 \beta_4 - \beta_3^2}{\beta_2^3} =  \frac{1490795}{1016064} + \frac{279\, \zeta (3)}{56}= 7.45604 \, .
\nonumber \\ 
\end{eqnarray}
\end{subequations}
From Eq.~(\ref{solx}) we obtain after a little algebra 
\begin{eqnarray}
\label{solu2}
 \bar{u} = \frac{1}{s} \, \exp \left( - \frac{\beta_3}{\beta_2} \, \frac{\ln s}{s} \left[  1  + O \left(  \frac{\ln^2 s}{s^2}, \frac{1}{s^2} \right) \right] \right)   \, .
\end{eqnarray}
Here, we have used the shorthand notation
\begin{eqnarray}
\label{defs}
s = \beta_2  \ln \left(   |\brm{x}| /x_0 \right) = \frac{7}{2}  \ln \left(   |\brm{x}| /x_0 \right)  
\end{eqnarray}
for the position dependence.

Now, we solve the remaining characteristics. It is to our advantage that the flow equations (\ref{charTau}) to (\ref{charZ}) and (\ref{multichar}) are all of the same structure. Thus, we can treat them simultaneously, by solving
\begin{eqnarray}
\label{Q}
\ell \frac{\partial \ln \bar{\chi}(\ell)}{\partial \ell}  = \chi_1 \, \bar{u} + \chi_2 \, \bar{u}^2 + O \left( \bar{u}^3 \right) \, ,
\end{eqnarray}
where $\bar{\chi}$ is a wildcard for $\bar{\tau}$, $\bar{w}_r$ and $\bar{Z}$. $\chi_0$ and $\chi_1$ are wildcards for the corresponding coefficients featured in (\ref{wilson2loop}) and (\ref{multiGamma}). Using $\ell \partial /\partial \ell = \beta \partial /\partial  \bar{u}$ and separating variables we obtain
\begin{eqnarray}
\frac{d \, \bar{\chi}}{\bar{\chi}} = \frac{\chi_1}{\beta_2} \frac{d \bar{u}}{\bar{u}} + \frac{\beta_2 \chi_2 - \beta_3 \chi_1}{\beta_2^2} \, d\bar{u} + O \left( \bar{u} \right) d\bar{u}  \, .
\end{eqnarray}
Now, integration is straightforward. By exponentiating the result we get
\begin{eqnarray}
\label{resQ}
\bar{\chi}( \bar{u})  &=& \chi_0 \, \bar{u}^{\chi_1/\beta_2} \exp \left( \frac{\beta_2 \chi_2 - \beta_3 \chi_1}{\beta_2^2} \,  \bar{u} \right)
\nonumber \\
&\times&  \left( 1+ O \left( \bar{u}^2 \right) \right)  \, ,
\end{eqnarray}
with $\chi_0$ being an integration constant.

\subsection{Final results -- logarithmic corrections}
After having solved the flow equations and computed the amplitudes, we are in the position to write down the critical behavior of the quantities of interest. 

\subsubsection{Percolation correlation function}
Our result for the percolation correlation function reads
\begin{eqnarray}
\label{resP1}
|\brm{x}|^4 \, \frac{P (\brm{x})}{P_0}  &=&  \left[ \bar{u}^{-1} + B_P  \right]^{a_P} \exp \left( c_P \, \bar{u}  \right) 
\nonumber \\
&\times& \left[ 1+ O \left( \bar{u}^2 \right) \right]  \, ,
\end{eqnarray}
where $P_0$ is a nonuniversal constant and 
\begin{subequations}
\begin{eqnarray}
a_P &=& - \frac{\gamma_1}{\beta_2} = \frac{1}{21} = 0.04762 \, ,
\\
c_P &=& \frac{\beta_2  \,\gamma_2 - \beta_3  \,\gamma_1}{\beta_2^2} = - \frac{103}{1323} 
\nonumber \\
&=& -0.07785 \, ,
\\
B_P &=& \frac{A_P (X_0)}{a_P} = \frac{7}{2} \left[ \frac{5}{6} +  \mathcal{Z} (X_0)\right] 
\nonumber \\
&=&  2.91667  + 3.5 \, \mathcal{Z} (X_0) \, .
\end{eqnarray}
\end{subequations}
Note that we have arranged things so that the 1-loop amplitude $A_P$ is not intermingled with the 2-loop contributions from the RG mapping. Equation~(\ref{resP1}) as it stands can be viewed as a parametric representation for the percolation correlation function. This result may be compared to simulations, e.g., by simply generating a parametric plot of $(|\brm{x}| ,  |\brm{x}|^4 P (\brm{x}))$ [cf.~Eq.~(\ref{solx})] and then comparing the numerical data to this plot. We can also cast our result in a more traditional form by using Eq.~(\ref{solu2}). Taylor expansion and a little algebra leads to
\begin{eqnarray}
\label{resP2}
|\brm{x}|^4 \, \frac{P (\brm{x})}{P_0}  &=&   \left[ s + B_P  \right]^{a_P} \bigg\{ 1 - \frac{b_P \ln s + c_P}{s} \nonumber \\
&+& O \left( \frac{\ln^2 s}{s^2}, \frac{\ln s}{s^2} , \frac{1}{s^2} \right) \bigg\} \, ,
\end{eqnarray}
where
\begin{eqnarray}
b_P = - \frac{\gamma_1 \beta_3}{\beta_2^2} = - \frac{671}{5292} = -0.12680 \, .
\end{eqnarray}

Eqs.~(\ref{resP1}) and (\ref{resP2}) show that the parametric representation is in comparison to the traditional form somewhat more systematic because it involves only one expansion variable, viz.~the effective coupling constant $\bar{u}$.€
In Eq.~(\ref{resP2}), on the other hand, functions of the position such as $1/s$, $\ln s/ s$, $\ln s/s^2$ and so on compete against each other and the ordering of the perturbation calculation is not so straightforward.

A closer look at (\ref{resP2}) and the definition~(\ref{defs}) of $s$ brings about following observation: by rescaling $x_0 \to X_0 \, x_0$ one can remove the explicit dependence of Eq.~(\ref{resP2}) on the arbitrary constant $X_0$ . Hence, our result on the percolation probability~(\ref{resP2}) features at minimum 2 fit parameters, viz.\ the length scale $x_0$ and the constant $P_0$. We choose, however, to keep $X_0$ in our formula because this way we have a further fit parameter at our command that can mimic higher order terms in the loop expansion. A likewise reasoning applies also to the other results that remain to be stated.

We would like to prevent the impression that one can remove the 1-loop amplitudes entirely from our results via rescaling. Of course one can eliminate one of the amplitudes, say $B_P$. For removing the amplitudes from several observables, however, one has to rescale $x_o$ individually at each attempt which leads to inconsistent results.

\subsubsection{Average resistance}
For the average resistance we obtain
\begin{eqnarray}
\label{resR1}
|\brm{x}|^{-2} \frac{M_R (\brm{x})}{M_{R,0}} &=&  \left[ \bar{u}^{-1} + B_R  \right]^{a_R} \exp \left( c_R \, \bar{u}  \right)  \left[ 1+ O \left( \bar{u}^2 \right) \right]
\nonumber \\
&=& \left[ s + B_R  \right]^{a_R} \bigg\{ 1 - \frac{b_R \ln s + c_R}{s} \nonumber \\
&+& O \left( \frac{\ln^2 s}{s^2}, \frac{\ln s}{s^2} , \frac{1}{s^2} \right) \bigg\} \, ,
\end{eqnarray}
with $M_{R,0}$ being a nonuniversal constant and where
\begin{subequations}
\begin{eqnarray}
a_R &=& - \frac{\zeta_{1,1}}{\beta_2} = - \frac{4}{21} = -0.19048 \, ,
\\
b_R &=& - \frac{\zeta_{1,1} \beta_3}{\beta_2^2} =  \frac{671}{1323} = 0.50718 \, ,
\\
c_R &=& \frac{\beta_2  \,\zeta_{1,2} - \beta_3  \,\zeta_{1,1}}{\beta_2^2} =  \frac{355}{2646} = 0.13417 \, ,
\\
B_R &=& \frac{A_{R_1} (X_0)}{a_R} = \frac{7}{2} \left[ \frac{11}{24} +  \mathcal{Z} (X_0)\right]   
\nonumber \\
&=& 1.60417 + 3.5 \, \mathcal{Z} (X_0) \, .
\end{eqnarray}
\end{subequations}

\subsubsection{Fractal masses}
Our results for the fractal mass of the backbone is
\begin{eqnarray}
\label{resMB}
|\brm{x}|^{-2} \frac{M_B (\brm{x})}{M_{B,0}} &=&  \left[ \bar{u}^{-1} + B_B  \right]^{a_B} \exp \left( c_B \, \bar{u}  \right)  \left[ 1+ O \left( \bar{u}^2 \right) \right]
\nonumber \\
&=& \left[ s + B_B  \right]^{a_B} \bigg\{ 1 - \frac{b_B \ln s + c_B}{s} \nonumber \\
&+& O \left( \frac{\ln^2 s}{s^2}, \frac{\ln s}{s^2} , \frac{1}{s^2} \right) \bigg\} \, ,
\end{eqnarray}
with the coefficients and the amplitude
\begin{subequations}
\begin{eqnarray}
a_B &=& - \frac{\zeta_{-1,1}}{\beta_2} =  \frac{1}{21} = 0.04762 \, ,
\\
b_B &=& - \frac{\zeta_{-1,1} \beta_3}{\beta_2^2} =  -\frac{671}{5292} = -0.12680 \, ,
\\
c_B &=& \frac{\beta_2  \,\zeta_{-1,2} - \beta_3  \,\zeta_{-1,1}}{\beta_2^2} =  \frac{86}{1323} 
\nonumber \\
&=& 0.06500 \, ,
\\
B_B &=& \frac{A_{R_{-1}} (X_0)}{a_B} = \frac{7}{2} \left[ \frac{5}{6} +  \mathcal{Z} (X_0)\right]   
\nonumber \\
&=& 2.91667 + 3.5 \, \mathcal{Z} (X_0) \, ,
\end{eqnarray}
\end{subequations}
as well as the nonuniversal constant $M_{B,0}$. 

For the mass of the red bonds we obtain
\begin{eqnarray}
\label{resMred}
|\brm{x}|^{-2} \frac{M_{\text{red}} (\brm{x})}{M_{\text{red},0}} &=&  \left[ \bar{u}^{-1} + B_{\text{red}}  \right]^{a_{\text{red}}} \exp \left( c_{\text{red}} \, \bar{u}  \right)  
\nonumber \\
&\times& \left[ 1+ O \left( \bar{u}^2 \right) \right]
\nonumber \\
&=& \left[ s + B_{\text{red}} \right]^{a_{\text{red}}} \bigg\{ 1 - \frac{b_{\text{red}} \ln s + c_{\text{red}}}{s} \nonumber \\
&+& O \left( \frac{\ln^2 s}{s^2}, \frac{\ln s}{s^2} , \frac{1}{s^2} \right) \bigg\} \, ,
\end{eqnarray}
where $M_{\text{red},0}$ 
\begin{subequations}
\begin{eqnarray}
a_{\text{red}}&=& - \frac{\zeta_{\infty,1}}{\beta_2} =  -\frac{5}{21} = -0.23810 \, ,
\\
b_{\text{red}} &=& - \frac{\zeta_{\infty,1} \beta_3}{\beta_2^2} =  \frac{3355}{5292} 
= 0.63398 \, ,
\\
c_{\text{red}} &=& \frac{\beta_2  \,\zeta_{\infty,2} - \beta_3  \,\zeta_{\infty,1}}{\beta_2^2} =  \frac{653}{5292} 
\nonumber \\
&=& 1.12339 \, ,
\\
B_{\text{red}} &=& \frac{A_{R_{\infty}} (X_0)}{a_{\text{red}}} = \frac{7}{2} \left[ \frac{13}{30} +  \mathcal{Z} (X_0)\right]   
\nonumber \\
&=& 1.51667 + 3.5 \, \mathcal{Z} (X_0) 
\end{eqnarray}
\end{subequations}
and where $M_{\text{red},0}$ is a nonuniversal constant.

The mass of the chemical path behaves in 6 dimensions according to
\begin{eqnarray}
\label{resMmin}
|\brm{x}|^{-2} \frac{M_{\text{min}} (\brm{x})}{M_{\text{min},0}} &=&  \left[ \bar{u}^{-1} + B_{\text{min}}  \right]^{a_{\text{min}}} \exp \left( c_{\text{min}} \, \bar{u}  \right)  
\nonumber \\
&\times& \left[ 1+ O \left( \bar{u}^2 \right) \right]
\nonumber \\
&=& \left[ s + B_{\text{red}} \right]^{a_{\text{min}}} \bigg\{ 1 - \frac{b_{\text{min}} \ln s + c_{\text{min}}}{s} \nonumber \\
&+& O \left( \frac{\ln^2 s}{s^2}, \frac{\ln s}{s^2} , \frac{1}{s^2} \right) \bigg\} \, ,
\end{eqnarray}
where $M_{\text{min},0}$ is, of course, a nonuniversal constant and
\begin{subequations}
\begin{eqnarray}
a_{\text{min}}&=& - \frac{\zeta_{0,1}}{\beta_2} =  -\frac{1}{6} = -0.16667 \, ,
\\
b_{\text{min}} &=& - \frac{\zeta_{0,1} \beta_3}{\beta_2^2} =  \frac{671}{1512} = 0.44378 \, ,
\\
c_{\text{min}} &=& \frac{\beta_2  \,\zeta_{0,2} - \beta_3  \,\zeta_{0,1}}{\beta_2^2} =  \frac{937}{6048} + \frac{5\, \ln 2}{56} -  \frac{9\, \ln 3}{112} 
\nonumber \\
&=& 0.12853 \, ,
\\
B_{\text{min}} &=& \frac{A_{R_{0}} (X_0)}{a_{\text{min}}} = \frac{7}{2} \left[ \frac{4}{21} + \frac{3\, \ln 2}{7} + \mathcal{Z} (X_0)\right]   
\nonumber \\
&=& 1.70639 + 3.5 \, \mathcal{Z} (X_0) \, .
\end{eqnarray}
\end{subequations}

\subsubsection{Multifractal moments}
Our result for the multifractal moments remains to be stated. We find
\begin{eqnarray}
\label{resMIl}
|\brm{x}|^{-2} \frac{ M_I^{(l)} (\brm{x}) }{M_{I,0}^{(l)}} &=&  \left[ \bar{u}^{-1} + B_I^{(l)}  \right]^{a_I^{(l)}} \exp \left( c_I^{(l)} \, \bar{u}  \right)  \left[ 1+ O \left( \bar{u}^2 \right) \right]
\nonumber \\
&=& \left[ s + B_I^{(l)}  \right]^{a_I^{(l)}} \bigg\{ 1 - \frac{b_I^{(l)} \ln s + c_I^{(l)}}{s} \nonumber \\
&+& O \left( \frac{\ln^2 s}{s^2}, \frac{\ln s}{s^2} , \frac{1}{s^2} \right) \bigg\} \, ,
\end{eqnarray}
with nonuniversal constants $M_{I,0}^{(l)}$ and where  Eq.~(\ref{resMIl}) are given by
\begin{subequations}
\label{multcoeffs}
\begin{eqnarray}
a_I^{(l)} &=& - \frac{\gamma_1^{(l)}}{\beta_2}  \, ,
\\
b_I^{(l)} &=& - \frac{\gamma_1^{(l)} \beta_3}{\beta_2^2}   \, ,
\\
c_I^{(l)} &=& \frac{\beta_2  \,\gamma_2^{(l)} - \beta_3  \,\gamma_1^{(l)}}{\beta_2^2}  \, ,
\\
B_I^{(l)} &=& \frac{A_I^{(l)} (X_0)}{a_I^{(l)}}\, .
\end{eqnarray}
\end{subequations}
The final formulae for the coefficients and the amplitude as a function of $l$ are somewhat lengthy, in particular those for $c_I^{(l)}$ and $B_I^{(l)}$. Therefore, we refrain from stating them explicitly and rather list the corresponding numerical values for $l = 0, \cdots, 5$ in Table~\ref{tab:numvals}. Note that the values for $l=0$ and $l=1$ coincide with those for the backbone and the average resistance, respectively. Moreover, for $l\to \infty$ the values for the red bonds are approached. Hence, our results satisfy the important consistency checks $M_I^{(0)} \sim M_B$, $M_I^{(l)} \sim M_R$ (since $M_R = C^{(1)}_R$), and $\lim_{l \to \infty} M_I^{(l)} \sim M_{\text{red}}$.
\begin{table}
\caption{The coefficients $a_I^{(l)}$, $b_I^{(l)}$, and $c_I^{(l)}$ as well as the amplitude $B_I^{(l)}$  appearing in Eq.~(\ref{multcoeffs}).}
\label{tab:numvals}
\begin{tabular}{c||c|c|c|c}
\hline \hline
$\quad l \quad $ & $a_I^{(l)}$ & $b_I^{(l)}$ & $c_I^{(l)}$ & $B_I^{(l)}-3.5\, \mathcal{Z} (X_0)$ \\ 
\hline
$ 0$ & $0.04762$ & $-0.12680$ & $0.06500$ & $2.91667$ \\
\hline
$ 1 $ & $-0.19048$ & $0.50718$ & $0.13417$ & $1.60417$ \\
\hline
$ 2 $ & $-0.21905$ & $0.58326$ & $0.13345$ & $1.52428$ \\
\hline
$ 3 $ & $-0.22789$ & $0.60681$ & $0.13150$ & $1.50734$ \\
\hline
$ 4 $ & $-0.23175$ & $0.61707$ & $0.12989$ & $1.50356$ \\
\hline
$ 5 $ & $-0.23377$ & $0.62245$ & $0.12868$ & $1.50335$ \\
\hline
$\to \infty$ & $-0.23810$ & $0.63398$ & $0.12339$ & $1.51667$ \\
\hline \hline
\end{tabular}
\end{table}

\section{Concluding remarks}
\label{concusions} 
We have determined the critical behavior of various geometrical and transport properties of percolation at the upper critical dimension $d=6$. Our investigation comprised the percolation correlation function, the fractal masses of the backbone, the red bonds and the shortest path as well as the multifractal moments of the current distribution. To our knowledge, the logarithmic corrections to these quantities have not been determined so far, not even to leading order. Our results are new and do not just represent a refinement of previous results.

Our analysis presented here benefited substantially from two concepts we introduced earlier, namely our real-world interpretation of Feynman diagrams and our notion of master operators. The real-world interpretation makes the abstract replicated field theory of RRN more intuitive and it provides practical guidance in calculations. The concept of master operators simplifies the analysis of dangerous irrelevant operators tremendously, because one is spared the computation and diagonalization of giant renormalization matrices.

The results presented in this paper satisfy several consistency checks. We verified that $M_I^{(l)} \sim M_R$ as it should since $M_R = C^{(1)}_R$. Furthermore, our results are reassured by satisfying  $M_I^{(0)} \sim M_B$ and $\lim_{l \to \infty} M_I^{(l)} \sim M_{\text{red}}$.

Given the computer hardware and sophisticated algorithms available today, our results should be testable by numerical simulations. Because we went beyond just calculating the leading corrections, we expect our results to compare well with simulations, perhaps even quantitatively. We hope that corresponding numerical work will be carried out in the near future. 

\begin{acknowledgments}
This work has been supported by the Deutsche Forschungsgemeinschaft via the Sonderforschungsbereich~237 ``Unordnung und gro{\ss}e Fluktuationen'' and the Emmy Noether-Programm. We thank D. Stauffer for bringing several references to our attention.

\end{acknowledgments} 

\appendix
\section{Amplitudes}
\label{app:amplitudes}
\begin{figure}
\epsfxsize=8.4cm
\epsffile{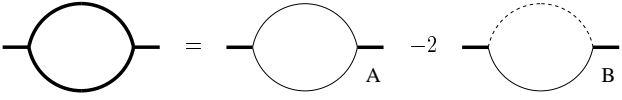}
\caption[]{\label{decomposition}2-leg Feynman diagrams for the RRN to two-loop order. The diagrams are assembled from the 3-leg vertex $-g$ and the bold propagator $G^{\text{bold}} ( \brm{p}, \vec{\lambda} ) = G ( \brm{k}, \vec{\lambda} ) \{ 1 - \delta_{\vec{\lambda}, \vec{0}} \}$,where $G (\brm{p}, \vec{\lambda}) = [\tau + \brm{p}^2 - w_r \Lambda_r (\vec{\lambda})]^{-1}$. Due to the factor $\{ 1 - \delta_{\vec{\lambda}, \vec{0}} \}$, which enforces the constraint $\psi ({\bf x},\vec{\lambda} = \vec{0}) =0$, the bold propagator decomposes in a conducting part $G^{\text{cond}} ( \brm{p}, \vec{\lambda}) = G ( \brm{p}, \vec{\lambda})$ carrying replica currents and an insulating part $G^{\text{ins}} ( \brm{p}) = G ( \brm{p}, \vec{\lambda})\delta_{\vec{\lambda}, \vec{0}}$ not carrying replica currents. Hence the bold diagram decomposes into the conducting diagrams A and B. The bold lines symbolize bold propagators, the light lines stand for conducting and the dashed lines for insulating propagators.}
\end{figure}
In this Appendix we outline the computation of the amplitudes entering the logarithmic corrections. 

\subsection{The linear/nonlinear RRN}
As a prerequisite, we need to know the 2-point correlation function, as a function of the position space coordinate $\brm{x}$, at 0-loop level. Hence we have to calculate
\begin{eqnarray}
\label{01}
G_2^{(0)} (\brm{x}, \vec{\lambda}) = \int_{\brm{p}} \frac{\exp (i \brm{p} \cdot \brm{x})}{\tau + \brm{p}^2 - w_r \Lambda_r (\vec{\lambda})} \, ,
\end{eqnarray}
where $\int_{\brm{p}}$ is an abbreviation for $1/(2 \pi)^{d/2} \int d^d p$. Employing Schwinger representation, we recast (\ref{01}) as
\begin{eqnarray}
\label{02}
G_2^{(0)} (\brm{x}, \vec{\lambda}) &=& \int_0^\infty ds \, \exp (-s \tau + s w_r \Lambda_r (\vec{\lambda})) 
\nonumber \\
&\times& \int_{\brm{p}} \exp (i \brm{p} \cdot \brm{x} - s \, \brm{p}^2) \, .
\end{eqnarray}
Completing the square in the exponential renders the momentum integration straightforward. After expanding to linear order in $w_r$ we get
\begin{eqnarray}
\label{03}
&&G_2^{(0)} (\brm{x}, \vec{\lambda}) = \frac{1}{(4 \pi)^{d/2}} \int_0^\infty ds \, \exp \left(-s \tau - \frac{\brm{x}^2}{4 \, s} \right)
\nonumber \\
&&\times\, \left\{  s^{-d/2}  + w_r \Lambda_r (\vec{\lambda}) s^{1-d/2} \right\} 
\nonumber \\
&&=\, \frac{1}{(2 \pi)^{d/2}} \, \left(  \frac{\sqrt{\tau}}{|\brm{x}|} \right)^{\frac{d-2}{2}} K_{\frac{d-2}{2}} (\sqrt{\tau}|\brm{x}|) 
\nonumber \\
&&+\,    \frac{w_r \Lambda_r (\vec{\lambda})}{2\, (2 \pi)^{d/2}} \, \left(  \frac{\sqrt{\tau}}{|\brm{x}|} \right)^{\frac{d-4}{2}} K_{\frac{d-4}{2}} (\sqrt{\tau}|\brm{x}|) \, ,
\end{eqnarray}
where $K_\nu (z)$ stands for the modified Bessel function~\cite{abramowitz_stegun_65}. We are interested in criticality. For vanishing $\tau$, Eq.~(\ref{03}) reduces to
\begin{eqnarray}
\label{04}
&&G_2^{(0)} (\brm{x}, \vec{\lambda}) = \frac{\Gamma (2 - \varepsilon/2)}{(4 \pi)^{d/2}} \left(  \frac{\brm{x}^2}{4} \right)^{-2 + \varepsilon/2}
\nonumber \\
&&+\,  w_r \Lambda_r (\vec{\lambda}) \, \frac{\Gamma (1 - \varepsilon/2)}{(4 \pi)^{d/2}} \left(  \frac{\brm{x}^2}{4} \right)^{-1 + \varepsilon/2} \, .
\end{eqnarray}
Below we will use the abbreviated notion
\begin{eqnarray}
\label{05}
G_2^{(0)} (\brm{x}, \vec{\lambda}) =  G_2^{(0)} (\brm{x})+ w_r \Lambda_r (\vec{\lambda}) \, G_2^{\prime (0)} (\brm{x}) \, .
\end{eqnarray}

Now we turn to 1-loop order. It should be clear from Sec.~\ref{renormalization} that the amplitudes entering the logarithmic corrections pertain to correlation functions and not vertex functions. Hence, we have to compute Feynman diagrams with their external legs attached and not amputated. Diagram A as displayed in Fig.~\ref{decomposition} stands for 
\begin{eqnarray}
&&\mbox{A} = \frac{g^2}{2} \int_{\brm{p}} \frac{\exp (i \brm{p} \cdot \brm{x})}{[\brm{p}^2 - w_r \Lambda_r (\vec{\lambda})]^2} 
 \\
&& \times \,  \int_{\brm{k}} \sum_{\vec{\kappa}} \, \frac{1}{\brm{k}^2 - w_r \Lambda_r (\vec{\kappa})}  \, 
\frac{1}{(\brm{k} + \brm{p})^2 - w_r \Lambda_r (\vec{\kappa} + \vec{\lambda} )} \, ,
\nonumber
\end{eqnarray}
where we have set $\tau =0$. We find it convenient to use Schwinger representation for the further steps. In this representation the integration over the loop momentum $\brm{k}$ is straightforward after  completing a square. The summation over the loop current $\vec{\kappa}$ is not so easy because it is not of Gaussian type for general $r$. One has to resort to the saddle-point approximation. Using our real-world interpretation, however, solving the saddle-point equation reduces to determining the total resistance of a diagram with its external legs amputated. We obtain
\begin{eqnarray}
&&\mbox{A} = \frac{g^2}{2} \, \frac{1}{(4 \pi)^{d/2}} \, \int_{\brm{p}} \exp (i \brm{p} \cdot \brm{x}) \int_0^\infty ds_1 ds_2 ds_3 
\nonumber \\
&& \times \, \frac{s_3}{(s_1 + s_2)^{d/2}} \exp \Big[  s_3 w_r \Lambda_r (\vec{\lambda}) + R_r (s_1 , s_2) w_r \Lambda_r (\vec{\lambda}) 
\nonumber \\
&&
- \, \frac{s_1 \, s_2}{s_1 + s_2} \brm{p}^2\Big] \, ,
\end{eqnarray}
where $R_r (s_1 , s_2)$ is the total nonlinear resistance of diagram A without external legs. For $r=1$ diagram A behaves like an Ohmic network. Hence $R_1 (s_1 , s_2) = s_1 s_2 /(s_1 + s_2)$. In the limit $r\to -1^+$, the total resistance of the diagram without external legs is nothing but the sum of the Schwinger parameters of the internal conducting propagators, i.e., $R_{-1} (s_1 , s_2) = s_1 + s_2$. For $r \to \infty$, blobs (multiple connections) of conducting propagators do not contribute and hence $R_{\infty} (s_1 , s_2) = 0$. On the case $r\to 0^+$ we will elaborate further below.

To carry out the remaining momentum integration, we once more complete a square. After expansion to linear order in $w_r$ we have
\begin{eqnarray}
\label{A2}
\mbox{A} = \frac{g^2}{2} \,  \left\{ \Pi_1 + w_r \Lambda_r (\vec{\lambda}) \left[  \Pi_2 + \Pi (r) \right] \right\} \, .
\end{eqnarray}
Here we have introduced abbreviations for the following integrals over Schwinger parameters
\begin{eqnarray}
\label{Pi}
\Pi_1 &=&  \frac{1}{(4 \pi)^{d}} \, \int_0^\infty ds_1 ds_2 ds_3 \, \frac{s_3}{[s_1 s_2 + s_1 s_3 + s_2 s_3]^{d/2}} 
\nonumber \\ &\times&
 \exp \left[ -  \frac{s_1 + s_2}{s_1 s_2 + s_1 s_3 + s_2 s_3} \, \frac{\brm{x}^2}{4} \right] \, ,
\\
\Pi_2 &=&  \frac{1}{(4 \pi)^{d}} \, \int_0^\infty ds_1 ds_2 ds_3 \, \frac{s_3^2}{[s_1 s_2 + s_1 s_3 + s_2 s_3]^{d/2}} 
\nonumber \\ &\times&
 \exp \left[ -  \frac{s_1 + s_2}{s_1 s_2 + s_1 s_3 + s_2 s_3} \, \frac{\brm{x}^2}{4} \right] \, ,
\\
\Pi (r) &=&  \frac{1}{(4 \pi)^{d}} \, \int_0^\infty ds_1 ds_2 ds_3  \, \frac{s_3 \, R_r (s_1 , s_2)}{[s_1 s_2 + s_1 s_3 + s_2 s_3]^{d/2}} 
\nonumber \\ &\times&
 \exp \left[ -  \frac{s_1 + s_2}{s_1 s_2 + s_1 s_3 + s_2 s_3} \, \frac{\brm{x}^2}{4} \right] \, .
\end{eqnarray}
Examples for the computation of these integrals as well as a list of results are given in Appendix~\ref{schwinger}.

The computation of diagram B is comparatively simple because it does not involve a summation over a loop current. We obtain
\begin{eqnarray}
\label{B1}
\mbox{B} = \frac{g^2}{2} \,  \left\{ \Pi_1 + w_r \Lambda_r (\vec{\lambda}) \left[  \Pi_2 + \Pi_3 \right] \right\} \, ,
\end{eqnarray}
where
\begin{eqnarray}
\label{Pi3}
\Pi_3 &=&  \frac{1}{(4 \pi)^{d}} \, \int_0^\infty ds_1 ds_2 ds_3 \, \frac{s_3 \, s_1}{[s_1 s_2 + s_1 s_3 + s_2 s_3]^{d/2}} 
\nonumber \\ &\times&
 \exp \left[ -  \frac{s_1 + s_2}{s_1 s_2 + s_1 s_3 + s_2 s_3} \, \frac{\brm{x}^2}{4} \right] \, .
\end{eqnarray}
Note that $s_1$ is nothing but the nonlinear total resistance of diagram B without external legs.

\subsubsection{$r = 1$}
Gathering the 0-loop result and the results for diagrams A and B we obtain
\begin{eqnarray}
\label{11}
&&G_2 (\brm{x}, \vec{\lambda}) =  G_2^{(0)} (\brm{x}) \left\{  1 + g^2 G_\varepsilon \left(  \frac{|\brm{x}|}{2} \right)^{\varepsilon}  \left[  \frac{1}{6\, \varepsilon} + \frac{5}{36} + \frac{\gamma}{6} \right] \right\}  
\nonumber \\ &&
- \, w \vec{\lambda}^2 \, G_2^{\prime (0)} (\brm{x}) \left\{  1 - g^2 G_\varepsilon \left(  \frac{|\brm{x}|}{2} \right)^{\varepsilon}  \left[  \frac{1}{2\, \varepsilon} + \frac{1}{6} + \frac{\gamma}{2} \right]  \right\} \, .
\end{eqnarray}
Next, we remove the $\varepsilon$ poles by employing our renormalization scheme~(\ref{renScheme}). The fact that the 1-loop renormalization factors $Z = 1 + \frac{u}{6 \, \varepsilon} + O (u^2)$ and $Z_{w_1} = 1 + \frac{5 \, u}{6 \, \varepsilon} + O (u^2)$ do indeed remove the $\varepsilon$ poles from the correlation function~(\ref{11}) represents an important consistency check for our calculation. Recalling our choice for the flow parameter $\ell$ we can write the renormalized correlation function as
\begin{eqnarray}
\label{12}
&&G_2 (\brm{x}, \vec{\lambda}) =  G_2^{(0)} (\brm{x}) \left\{  1 + u  \left[  \frac{5}{36} + \frac{\gamma}{6} + \frac{\ln X_0}{6}\right] \right\}  
\nonumber \\ & &
- \, w \vec{\lambda}^2 \, G_2^{\prime (0)} (\brm{x}) \left\{  1 - u  \left[   \frac{1}{6} + \frac{\gamma}{2} + \frac{\ln X_0}{2} \right]  \right\} \, .
\end{eqnarray} 
From Eq.~(\ref{12}) we can simply read off the amplitude $A_P (X_0)$. The result is stated in Eq.~(\ref{resAP}). Also, we can read off $A_{w_1}$. Using Eq.~(\ref{defARr}) we get the result for $A_{R_1}$ as stated in Eq.~(\ref{resAR1}).

\subsubsection{$r\to -1^+$}
Our 1-loop calculation leads to
\begin{eqnarray}
\label{-11}
G_2^\prime (\brm{x}) =  G_2^{\prime (0)} (\brm{x}) \left\{  1 + g^2 G_\varepsilon \left(  \frac{|\brm{x}|}{2} \right)^{\varepsilon}  \left[  \frac{1}{3\, \varepsilon} + \frac{5}{18} + \frac{\gamma}{3} \right] \right\} \, .
\nonumber \\ 
\end{eqnarray}
Upon renormalization we obtain
\begin{eqnarray}
\label{-12}
G_2^\prime (\brm{x}) =  G_2^{\prime (0)} (\brm{x}) \left\{  1 + u  \left[  \frac{5}{18} + \frac{\gamma}{3} + \frac{\ln X_0}{3} \right] \right\} \, .
\end{eqnarray}
Here we used the 1-loop result $Z_{w_{-1}} = 1 + O (u^2)$. Utilizing Eq.~(\ref{defARr}) we get our final result for $A_{R_{-1}}$, see Eq.~(\ref{resAR-1}).

\subsubsection{$r\to \infty$}
In the limit $r\to \infty$ we find
\begin{eqnarray}
\label{infty1}
G_2^\prime (\brm{x}) =  G_2^{\prime (0)} (\brm{x}) \left\{  1 - g^2 G_\varepsilon \left(  \frac{|\brm{x}|}{2} \right)^{\varepsilon}  \left[  \frac{2}{3\, \varepsilon} + \frac{2}{9} + \frac{2\, \gamma}{3} \right] \right\} \, .
\nonumber \\ 
\end{eqnarray}
Using $Z_{w_\infty} = 1 + \frac{u}{\varepsilon} + O (u^2)$ we obtain the renormalized version
\begin{eqnarray}
\label{infty2}
G_2^\prime (\brm{x}) =  G_2^{\prime (0)} (\brm{x}) \left\{  1 - u  \left[  \frac{2}{9} + \frac{2\, \gamma}{3} + \frac{2 \, \ln X_0}{3} \right] \right\} \, .
\end{eqnarray}
Exploiting Eq.~(\ref{defARr}) yields Eq.~(\ref{resAred}).

\subsubsection{$r\to 0^+$}
In the limit $r\to 0^+$ the diagrammatic resistance $R_{r} (s_1 , s_2)$ is determined by the shortest self avoiding path through the diagram with amputated legs, i.e., $R_{0} (s_1 , s_2) =  \min (s_1, s_2)$. We find it useful to write diagram A as
\begin{eqnarray}
\mbox{A} &=&  \frac{g^2}{2} \, \frac{1}{(4 \pi)^{d}} \, \int_0^\infty ds_1 ds_2 ds_3 \, \frac{s_3}{[s_1 s_2 + s_1 s_3 + s_2 s_3]^{d/2}} 
\nonumber \\ &\times&
 \exp \left[ -  \frac{s_1 + s_2}{s_1 s_2 + s_1 s_3 + s_2 s_3} \, \frac{\brm{x}^2}{4} \right] 
\nonumber \\ &\times&
\theta (s_2 - s_1) \exp \left[ -  (s_1 + s_3) i w_{0} \lambda \right] \, ,
\end{eqnarray}
where $\theta$ stands for the step function and $\lambda = \sum_{\alpha =1}^D \lambda^{(\alpha )}$. By virtue of $\theta (s_2 - s_1) -1 = - \theta (s_1 - s_2)$ it is convenient to treat diagrams A and B in one go. Expanding to linear order in $w_{0}$ we get
\begin{eqnarray}
\mbox{A} - 2 \, \mbox{B} &=&  i w_{0} \lambda \, g^2 \, \Pi_4  \, ,
\end{eqnarray}
where we dropped contributions independent of $\lambda$ for notational simplicity and where
\begin{eqnarray}
\label{Pi4}
&&\Pi_4 =  \frac{1}{(4 \pi)^{d}} \, \int_0^\infty ds_1 ds_2 ds_3 \, \frac{s_3 \, (s_1 + s_3)}{[s_1 s_2 + s_1 s_3 + s_2 s_3]^{d/2}} 
\nonumber \\ &&
\times \, \theta (s_1 - s_2) \,  \exp \left[ -  \frac{s_1 + s_2}{s_1 s_2 + s_1 s_3 + s_2 s_3} \, \frac{\brm{x}^2}{4} \right]  \, .
\end{eqnarray}
With the result for $\Pi_4$ from Appendix~\ref{schwinger} we find
\begin{eqnarray}
\label{min1}
G_2^\prime (\brm{x}) &=&  G_2^{\prime (0)} (\brm{x}) \bigg\{  1 - g^2 G_\varepsilon \left(  \frac{|\brm{x}|}{2} \right)^{\varepsilon}  
\nonumber \\
&\times&\left[  \frac{5}{12\, \varepsilon} - \frac{1}{36} + \frac{\ln 2}{4} + \frac{5\, \gamma}{12} \right] \bigg\} \, .
\end{eqnarray}
Upon renormalization, for which we here need $Z_{w_0} = 1 + \frac{3 \, u}{4 \, \varepsilon} + O (u^2)$, we get
\begin{eqnarray}
\label{min2}
G_2^\prime (\brm{x}) &=&  G_2^{\prime (0)} (\brm{x}) \bigg\{  1 + u \left[  \frac{1}{36} - \frac{\ln 2}{4}  -  \frac{5\, \gamma}{12}  - \frac{5\, \ln X_0}{12} \right] \bigg\} \, .
\nonumber \\ 
\end{eqnarray}
Equation~(\ref{defARr}) finally leads to $A_{R_{0}}$ as stated in Eq.~(\ref{resAmin}).

\subsection{The noisy RRN}
As above, we start by determining the 0-loop contribution. Without much afford we find
\begin{eqnarray}
G_2^{(0)} (\brm{x}, \tensor{\lambda})_{\mathcal{O}^{(l)}}  &=&  - v_l K_l (\tensor{\lambda}) \, \frac{\Gamma (1 - \varepsilon/2)}{(4 \pi)^{d/2}} \left(  \frac{\brm{x}^2}{4} \right)^{-1 + \varepsilon/2} 
\nonumber \\
&+& \cdots
\nonumber \\
&=& - v_l K_l (\tensor{\lambda}) \, G_2^{\prime (0)} (\brm{x}) + \cdots \, .
\end{eqnarray}

Now we turn to the 1-loop contributions. We have to compute the diagrams A and B with insertions of the operator $\mathcal{O}^{(l)}$. We once more employ our real world interpretation. Since we are interested here in the moments of the current distribution instead of the total resistance we now determine the moments of the current distribution for the diagrams rather than their resistance. For details on the method we refer to Refs.~\cite{stenull_janssen_epl_2000,stenull_janssen_2001}. We obtain for diagram A with $\mathcal{O}^{(l)}$ inserted  
\begin{eqnarray}
\mbox{A}_{\mathcal{O}^{(l)}} = - \frac{g^2}{2} \,  v_l K_l (\tensor{\lambda}) \, \left[  \Pi_2 + \Sigma (l) \right]  + \cdots \, ,
\end{eqnarray}
where 
\begin{eqnarray}
\Sigma (l) &=&  \frac{1}{(4 \pi)^{d}} \, \int_0^\infty ds_1 ds_2 ds_3  \, \frac{s_3 \, C^{(l)} (s_1 , s_2)}{[s_1 s_2 + s_1 s_3 + s_2 s_3]^{d/2}} 
\nonumber \\ &\times&
 \exp \left[ -  \frac{s_1 + s_2}{s_1 s_2 + s_1 s_3 + s_2 s_3} \, \frac{\brm{x}^2}{4} \right] \, .
\end{eqnarray}
The $\varepsilon$ expansion result for this integral can be found in Appendix~\ref{schwinger}. $C^{(l)} (s_1 , s_2)$ is the moment of the current distribution for diagram A without external legs,
\begin{eqnarray}
C^{(l)} (s_1 , s_2) = s_1 \left( \frac{s_2} {s_1 + s_2} \right)^{2l} + s_2 \left( \frac{s_1} {s_1 + s_2} \right)^{2l}  .
\end{eqnarray}
Diagram B with insertion can be written as
\begin{eqnarray}
\mbox{B}_{\mathcal{O}^{(l)}} = - \frac{g^2}{2} \,  v_l K_l (\tensor{\lambda}) \, \left[  \Pi_2 + \Pi_3 \right]  + \cdots \, ,
\end{eqnarray}
Note that $s_1$ is nothing but the moment of the current distribution for the diagram B with its external legs detached. 

Using the results of Appendix~\ref{schwinger} we obtain for the 2-point correlation function with insertion
\begin{eqnarray}
&&G_2 (\brm{x}, \vec{\lambda})_{\mathcal{O}^{(l)}} =  - v_l K_l (\tensor{\lambda}) \, G_2^{\prime (0)} (\brm{x}) \bigg\{  1 - g^2 G_\varepsilon \left(  \frac{|\brm{x}|}{2} \right)^{\varepsilon}  
\nonumber \\
&& \times \, \bigg[  \frac{2}{3\, \varepsilon} + \frac{2}{9} + \frac{2\, \gamma}{3} - \frac{1}{(2l+1)(2l+2)} \bigg( \frac{2}{\varepsilon} -1 + 2 \gamma 
\nonumber \\
&&- \, \Psi (2l+1) -\Psi (2) + 2 \Psi (2l+3 )\bigg) \bigg]  \bigg\} + \cdots \, .
\end{eqnarray}
Recalling our result for the renormalization factor $Z^{(l)}$ to 1-loop order,
\begin{eqnarray}
Z^{(l)} = 1  
-  \frac{ 1 - 15\,l - 10\,l^2 }{6\,\left( 1 + l \right) \,\left( 1 + 2\,l \right)}  \, \frac{u}{\varepsilon} + O(u^2) \, 
\end{eqnarray}
we find upon renormalization
\begin{eqnarray}
&&G_2 (\brm{x}, \vec{\lambda})_{\mathcal{O}^{(l)}} =  - v_l K_l (\tensor{\lambda}) \, G_2^{\prime (0)} (\brm{x}) \bigg\{  1 - u \bigg[  \frac{2}{9}
\nonumber \\
&& + \,  \frac{2}{3} \, \mathcal{Z} (X_0)  - \frac{1}{(2l+1)(2l+2)} \bigg( -1 + 2 \, \mathcal{Z} (X_0) 
\nonumber \\
&&- \, \Psi (2l+1) -\Psi (2) + 2 \Psi (2l+3 )\bigg) \bigg]  \bigg\} + \cdots  .
\end{eqnarray}
Via Eq.~(\ref{defAIl}) we finally get the amplitude stated in Eq.~(\ref{resAIl}).

\section{Integrals over Schwinger parameters}
\label{schwinger}
In this Appendix we sketch our computation of integrals introduced in Appendix~\ref{app:amplitudes}. Instead of elaborating on all the integrals we give two representative examples. Moreover, we give a comprehensive list of results.

\subsection{Examples}
As a first example we consider the integral $\Pi (1)$. We start manipulating it by setting $s_1 = t \, y$, $s_2 = t \, (1-y)$ and $s_3 = t \, z$. This change of variables yields
\begin{eqnarray}
\Pi (1) &=&  \frac{1}{(4 \pi)^{d}} \, \int_0^\infty dt dz \, \int_0^1 dy \, t^{4-d} \frac{zy(1-y)}{[z + y(1-y)]^{d/2}} 
\nonumber \\ &\times&
 \exp \left[ -  \frac{t^{-1}}{z + y(1-y)} \, \frac{\brm{x}^2}{4} \right] \, .
\end{eqnarray}
Next, we change variables so that the argument of the exponential function is simplified. To be specific, we switch from $t$ to the integration variable 
\begin{eqnarray}
t^\prime =  \frac{t^{-1}}{z + y(1-y)} \, \frac{\brm{x}^2}{4}  \, .
\end{eqnarray}
The integration over $t^\prime$ is straightforward and gives
\begin{eqnarray}
\Pi (1) &=&  \left( \frac{\brm{x}^2}{4} \right)^{5-d} \frac{\Gamma (d-5)}{(4 \pi)^{d}} \, \int_0^\infty dz \, \int_0^1 dy
\nonumber \\
&\times&   zy(1-y) \, [z + y(1-y)]^{d/2-5} \, .
\end{eqnarray}
This can now be simplified by switching from $z$ to $z^\prime = z/[y(1-y)]$. We obtain
\begin{eqnarray}
\Pi (1) &=&  \left( \frac{\brm{x}^2}{4} \right)^{5-d} \frac{\Gamma (d-5)}{(4 \pi)^{d}} \, B \left( \frac{d}{2} -1, \frac{d}{2} -1\right)
\nonumber \\
&\times&  \left\{ \frac{2}{6-d} -  \frac{2}{8-d} \right\} \, ,
\end{eqnarray}
where $B(n,m) = \Gamma (n) \Gamma (m) / \Gamma (n+m)$ is the Beta function~\cite{abramowitz_stegun_65}. Via expansion for small $\varepsilon =6-d$ and a little algebra we finally arrive at the result stated in Eq.~(\ref{resPi1}).

The second and last example we consider is $\Pi_4$. To get rid ot the step function we change variables by setting $s_1 = t_1 + t_2$, $s_2 = t_2$, and $s_3 = t_3$. This step yields
\begin{eqnarray}
\Pi_4 &=&  \frac{1}{(4 \pi)^{d}} \, \int_0^\infty dt_1 dt_2 dt_3 \, \frac{t_3 \, (t_1 + t_2 + t_3)}{[(t_1 + 2\,t_2) t_3 + (t_1 + t_2) t_2]^{d/2}} 
\nonumber \\
&\times&  \exp \left[ -  \frac{t_1 + 2\, t_2}{(t_1 + 2\,t_2) t_3 + (t_1 + t_2) t_2} \, \frac{\brm{x}^2}{4} \right]  \, .
\end{eqnarray}
$\Pi_4$ can be simplified further upon setting $t_1 = t \, y$, $t_2 = t \, (1-y)/2$ and $t_3 = t \, z$. we arrive at
\begin{eqnarray}
\Pi_4 &=&  \frac{1}{4 \, (4 \pi)^{d}}  \, \int_0^\infty dt dz \, \int_0^1 dy \, t^{4-d} \frac{z(1+y+2z)}{[z + \frac{1}{4}(1-y^2)]^{d/2}} 
\nonumber \\ &\times&
 \exp \left[ -  \frac{t^{-1}}{z + \frac{1}{4}(1-y^2)} \, \frac{\brm{x}^2}{4} \right] \, .
\end{eqnarray}
Next, the $t$ integration is rendered straightforward by going from $t$ to
\begin{eqnarray}
t^\prime =  \frac{t^{-1}}{z + \frac{1}{4}(1-y^2)} \, \frac{\brm{x}^2}{4}  \, .
\end{eqnarray}
Integrating out $t^\prime$ we get
\begin{eqnarray}
\Pi_4 &=&  \left( \frac{\brm{x}^2}{4} \right)^{5-d} \frac{\Gamma (d-5)}{4\, (4 \pi)^{d}} \, \int_0^\infty dz \, \int_0^1 dy
\nonumber \\
&\times&   z(1+y+2z) \, \left[z + \frac{1}{4}(1-y^2)\right]^{d/2-5} \, .
\end{eqnarray}
After simplifying the remaining integrations by introducing $z^\prime = 4z/(1-y^2)$ we obtain
\begin{eqnarray}
&&\Pi_4 =  \left( \frac{\brm{x}^2}{4} \right)^{5-d} \frac{\Gamma (d-5)}{(4 \pi)^{d}} \, 4^{1-d/2}
\nonumber \\
&&\times\,   \bigg\{ B \left( \frac{1}{2}, \frac{d}{2} -1 \right)  \bigg[ \frac{2}{4-d} - \frac{4}{6-d} +  \frac{2}{8-d} \bigg] 
\\
&&+\,    2   \left[ B \left( \frac{1}{2}, \frac{d}{2} -2 \right) + B \left( 1, \frac{d}{2} -2 \right)\right] \bigg[ \frac{2}{6-d} -  \frac{2}{8-d} \bigg] 
\bigg\}  .
\nonumber
\end{eqnarray}
$\varepsilon$ expansion and some rearrangements finally lead to the result~(\ref{resPi4}).

\subsection{Results}
Here we list our results for all the integrals over Schwinger parameters we used in calculating the 1-loop diagrams A and B:
\begin{eqnarray}
\Pi_1 &=& \left( \frac{\brm{x}^2}{4} \right)^{-2 + \varepsilon} \frac{\Gamma (2-\varepsilon/2)}{(4\, \pi)^{d/2}} \, G_\varepsilon \left\{ - \frac{1}{3\, \varepsilon} - \frac{5}{18} - \frac{\gamma}{3} \right\} ,
\nonumber \\
\\
\Pi_2 &=& \left( \frac{\brm{x}^2}{4} \right)^{-1 + \varepsilon} \frac{\Gamma (1-\varepsilon/2)}{(4\, \pi)^{d/2}} \, G_\varepsilon \left\{ - \frac{2}{3\, \varepsilon} - \frac{5}{9} - \frac{2\,\gamma}{3} \right\} ,
\nonumber \\
\\
\Pi_3 &=& \left( \frac{\brm{x}^2}{4} \right)^{-1 + \varepsilon} \frac{\Gamma (1-\varepsilon/2)}{(4\, \pi)^{d/2}} \, G_\varepsilon \left\{  \frac{1}{\varepsilon} + \frac{1}{2}  + \gamma \right\} ,
\\
\label{resPi1} 
\Pi (1) &=& \left( \frac{\brm{x}^2}{4} \right)^{-1 + \varepsilon} \frac{\Gamma (1-\varepsilon/2)}{(4\, \pi)^{d/2}} \, G_\varepsilon \left\{  \frac{1}{3\, \varepsilon} + \frac{1}{9} + \frac{\gamma}{3} \right\} ,
\nonumber \\
\Pi (-1) &=& 2 \, \Pi_3 \, ,
\\
\Pi (0) &=& 0 \, ,
\\
\label{resPi4}
\Pi_4 &=& \left( \frac{\brm{x}^2}{4} \right)^{-1 + \varepsilon} \frac{\Gamma (1-\varepsilon/2)}{(4\, \pi)^{d/2}} \, G_\varepsilon \bigg\{  \frac{5}{12\, \varepsilon} - \frac{1}{36}  
\nonumber \\
&+& \frac{\ln 2}{4} +\frac{5\,\gamma}{12} \bigg\} \, ,
\\
\Sigma (l) &=& \left( \frac{\brm{x}^2}{4} \right)^{-1 + \varepsilon} \frac{\Gamma (1-\varepsilon/2)}{(4\, \pi)^{d/2}} \, G_\varepsilon \, \frac{2}{(2l+1)(2l+2)} 
\nonumber \\ 
&\times&
\bigg\{  \frac{2}{3\, \varepsilon} - 1 + 2\,\gamma  - \Psi (2l+1) -\Psi (2) 
\nonumber \\
&& \hspace{3.5cm}+ 2 \Psi (2l+3 ) \bigg\} \, .
\end{eqnarray}
Note that these results fulfill several consistency checks, namely $\Sigma (1) = \Pi (1)$, $\Sigma (0) = \Pi (-1)$ and $\lim_{l\to \infty} \Sigma (l) = \Pi (0)$.


\end{document}